\documentclass[acmconf,authorversion]{acmart}

\renewcommand\footnotetextcopyrightpermission[1]{} 

\makeatletter
\renewcommand\@formatdoi[1]{\ignorespaces}
\makeatother

\newtheorem{theorem}{Definition}
\usepackage{cleveref}
\usepackage{multirow}

\AtBeginDocument{%
  \providecommand\BibTeX{{%
    \normalfont B\kern-0.5em{\scshape i\kern-0.25em b}\kern-0.8em\TeX}}}

\setcopyright{none}
\acmYear{2022}

\acmConference[CHI '22, TRAIT]{CHI '22, Workshop on Trust and Reliance in AI-Human Teams (trAIt)}{April 30, 2022}{New Orleans, LA, USA}
\acmBooktitle{CHI Conference on Human Factors in Computing Systems (CHI ’22), Workshop on Trust and Reliance in AI-Human Teams (trAIt), April 30, 2022, New Orleans, LA, USA}

\begin{document}

\title[Measuring Appropriate Reliance in Human-AI Decision-Making]{Should I Follow AI-based Advice? Measuring Appropriate Reliance in Human-AI Decision-Making}




\author{Max Schemmer}
\email{max.schemmer@kit.edu}
\affiliation{%
  \institution{Karlsruhe Institute of Technology}
  \city{Karlsruhe}
  \country{Germany}
}

\author{Patrick Hemmer}
\affiliation{%
  \institution{Karlsruhe Institute of Technology}
  \city{Karlsruhe}
  \country{Germany}
}
\email{patrick.hemmer@kit.edu}

\author{Niklas Kühl}
\affiliation{%
  \institution{Karlsruhe Institute of Technology}
  \city{Karlsruhe}
  \country{Germany}
}  
\email{niklas.kuehl@kit.edu}

\author{Carina Benz}
\affiliation{%
  \institution{Karlsruhe Institute of Technology}
  \city{Karlsruhe}
  \country{Germany}
}
\email{carina.benz@kit.edu}

\author{Gerhard Satzger}
\affiliation{%
  \institution{Karlsruhe Institute of Technology}
  \city{Karlsruhe}
  \country{Germany}
}
\email{gerhard.satzger@kit.edu}





\renewcommand{\shortauthors}{Schemmer et al.}

\begin{abstract}
Many important decisions in daily life are made with the help of advisors, e.g., decisions about medical treatments or financial investments. Whereas in the past, advice has often been received from human experts, friends, or family, advisors based on artificial intelligence (AI) have become more and more present nowadays. Typically, the advice generated by AI is judged by a human and either deemed reliable or rejected. However, recent work has shown that AI advice is not always beneficial, as humans have shown to be unable to ignore incorrect AI advice, essentially representing an over-reliance on AI. Therefore, the aspired goal should be to enable humans not to rely on AI advice blindly but rather to distinguish its quality and act upon it to make better decisions. Specifically, that means that humans should rely on the AI in the presence of correct advice and self-rely when confronted with incorrect advice, i.e., establish appropriate reliance (AR) on AI advice on a case-by-case basis. 
Current research lacks a metric for AR. This prevents a rigorous evaluation of factors impacting AR and hinders further development of human-AI decision-making. Therefore, based on the literature, we derive a measurement concept of AR. We propose to view AR as a two-dimensional construct that measures the ability to discriminate advice quality and behave accordingly. In this article, we derive the measurement concept, illustrate its application and outline potential future research.
\end{abstract}



\maketitle

\section{Introduction}
For many important decisions in life, we seek the opinion of advisors. While in the past, advice was typically obtained from human experts, nowadays, advisors based on artificial intelligence (AI) are becoming more and more present in research and practice \cite{jung2018robo}. For example, AI now advises medical professionals with regard to breast cancer screening \cite{mckinney2020international}, or in detecting COVID-19 pneumonia \cite{chowdhury2020can}. Past research has often focused on maximizing the utilization of advice \cite{sniezek2001trust,schultze2015effects,kuhlyou}, i.e., increasing the amount of accepted advice leading to increased compliance. Even though this might be valid for human advice seeking, a different paradigm might be more suitable in the scenario of human-AI decision-making due to a change in objectives. Research on human advice is often based on the perspective of the advisor, who is most likely interested in high advice utilization \cite{sniezek2001trust}. For example, when considering a bank that offers advice for investment decisions, the investments will have different potential outcomes for the bank itself. Therefore, the bank is interested in the client following its advice as it has ``stakes'' in the decision. However, the critical difference in the human-AI decision-making setting is that the advice seekers can design and develop the advisor based on personal goals. If those goals are to maximize decision-making performance, blindly following AI advice does not necessarily lead to the best possible outcome \cite{bansal2021does}. 
For this reason, researchers argue that in order to benefit the most from AI advice, humans need to be able to \textit{appropriately} rely on it \cite{bansal2021does,buccinca2021trust,zhang2020effect}. We define appropriate reliance (AR) in AI advice as the human's ability to differentiate between correct and incorrect AI advice and to act upon that discrimination. 

As AI becomes more important in our daily lives, both professionally and privately, humans need to be able to discriminate between correct and incorrect advice. If their discrimination capabilities are sufficient, they could even benefit from an AI that performs worse on average compared to them. However, the discriminating ability is not the only factor of appropriate reliance as humans need not only to discriminate but also to adapt their decisions accordingly. For example, in theory, they could be capable of detecting errors but do not dare to contradict AI advice due to a disproportionately high level of trust in or perceived authority/skill of the AI. 

To enable humans to rely on AI advice appropriately, we need to have a precise understanding of AR and measure it coherently. Currently, there is no specific measurement for AR with regard to AI advice, and researchers are using many different measurement concepts.  
Therefore, we derive a measurement concept for AR in the context of AI advice based on literature in automation on AR \cite{lee2004trust,talone2019effect} and organizational psychology \cite{sniezek2001trust}. Subsequently, we illustrate our measurement concept through a behavioral experiment. Lastly, we discuss possible avenues for future research.


The remainder of this article is structured as follows: In \Cref{sec:RW}, we first outline related work on AR in the context of human-AI decision-making. In \Cref{sec:concept}, we propose a measurement concept capturing AR, followed by an illustration drawn from a user study in \Cref{sec:illustration}. In \Cref{sec:discussion} we discuss AR and provide ideas for future work. \Cref{sec:conclusion} concludes our work.

\section{Related Work}\label{sec:RW}

Historically, many researchers have worked on AR with regard to automation \cite{lee2004trust} and robotics \cite{talone2019effect}. In the following, we will provide an overview of the most common definitions. Fundamental work in the context of AR in automation has been laid by \citet{lee2004trust}. The authors outline the relationship between ``appropriate trust'' and AR in their work. However, they do not define AR explicitly but provide examples of inappropriate reliance, such as ``misuse and disuse are two examples of inappropriate reliance on automation that can compromise safety and profitability'' \cite[p. 50]{lee2004trust}.
Other researchers go one step further and define inappropriate reliance as under- or over-reliance \cite{van2007aiding,yuviler2011effect,lu2021human}.

\citet{wang2008selecting} define appropriate reliance as the impact of reliance on performance. For example, they discuss the situation in which automation reaches a reliability of 99\%, and the human performance is 50\%. In their opinion, it would be appropriate to always rely on AI as this would increase performance. \citet{talone2019effect} follows the work by \citet{wang2008selecting} and defines AR as ``the pattern of reliance behavior(s) that is most likely to result in the best human-automation team performance'' \cite[p. 13]{talone2019effect}. Both see appropriate reliance as a function of team performance.

Recent work in human-AI decision-making has started to discuss AR in the context of AI advice. \citet{lai2021towards} gives an overview of empirical studies that analyze AI advice considering AR. For example, \citet{Chandrasekaran2018DoEM} analyze whether humans can learn to predict the model behavior. This ability is associated with an improved ability to rely on the model's predictions for the right cases. Moreover, \citet{gonzalez2020human} evaluate the impact of explainable AI (XAI) on the discrimination of incorrect and correct AI advice. Similarly, \citet[p. 1]{poursabzi2021manipulating} point out the idea of AR in the form of ``making people more closely follow a model’s predictions when it is beneficial for them to do so or enabling them to detect when a model has made a mistake''. However, the authors do not explicitly relate this idea to the concept of AR. In this context, additional work uses the term ``appropriate trust'' with a similar interpretation as the behavior to follow ``the fraction of tasks where participants used the model's prediction when the model was correct and did not use the model's prediction when the model was wrong'' \cite[p. 323]{wang2021explanations}. Finally, also \citet[p. 190]{yang2020visual} define ``appropriate trust is to [not] follow an [in]correct recommendation''. All these articles have in common that they consider AR or appropriate trust on a case-by-case basis. Similar to our work, \citet{buccinca2021trust} analyze whether cognitive forcing functions can reduce over-reliance on AI advice, which is measured as the percentage of agreement with the AI when the AI makes incorrect predictions. \citet{bussone2015role} assess how explanations impact trust and reliance on clinical decision support systems. The authors partition reliance into over- and self-reliance as part of their study. However, they use a qualitative approach to answer their research questions. To summarize, previous research does not provide a unified measurement concept that allows measuring AR on AI advice.

\section{Deriving a measurement concept of Appropriate Reliance on AI Advice}\label{sec:concept}
Despite several studies having examined human-AI interaction with regard to reliance, an agreed-upon definition of AR is still missing. We, therefore, initiate our research by deriving a definition of AR. To provide an accurate definition, we first analyze the two terms ``appropriate'' and ``reliance'' individually. 

\paragraph{Reliance.} Reliance itself is defined as a behavior \cite{lee2004trust,dzindolet2003role}. This means it is neither a feeling nor an attitude but the actual action conducted. Defining reliance as behavior also clarifies the role of trust, which is defined as ``the attitude that an agent will help achieve an individual’s goals in a situation characterized by uncertainty and vulnerability'' \cite[p. 51]{lee2004trust}. In general, research has shown that trust increases reliance, but it can also take place without trust being present \cite{lee2004trust}. For example, we might not trust the banking advisor but consciously decide that the advice is still the best possible decision. 
The final reliance is beyond trust and is also influenced by other attitudes such as perceived risk or self-confidence \cite{riley2018operator}.


\paragraph{Appropriateness.} After establishing a common understanding of reliance, we proceed by defining ``appropriateness''. The appropriateness of reliance stems from the fact that current AI is imperfect, i.e., it may provide erroneous advice. This erroneous advice can be divided into systematic errors and random errors \cite{talone2019effect}. While humans can identify systematic errors, random errors have no identifiable patterns and can not be distinguished. These different errors allow differentiation between two cases of AR. If all errors are random and cannot be detected, then humans should always rely on AI if, on average, AI performs better and never rely if AI performs worse on average \cite{talone2019effect}. 
However, suppose there are some systematic errors, depending on the discrimination capabilities. In that case, humans might be able to differentiate between correct and incorrect advice, which may even result in superior performance compared to the scenario of AI and humans conducting the task alone \cite{hemmer2021human}.
This changes the overall discrimination to a case-by-case discrimination. In the presence of systematic errors, humans should evaluate each case individually. Since the solution approach in the presence of just random errors is relatively simple, as pointed out above, in this article, we focus on the more complicated setting when a significant proportion of task instances inhibit systematic errors. 

For AR in the presence of systematic errors, we see two main aspects.
First, humans need to be able to differentiate between correct and incorrect advice. Second, people need to act upon their discrimination accordingly due to its behavioral nature. For instance, a human decision-maker might be able to differentiate between correct and incorrect AI advice but has a too high level of trust to reject the advice of the AI. Thus, AR consists of the capability to discriminate and execute the consequent behavior. 

\begin{theorem}
Appropriate reliance on AI advice is 
\begin{itemize}
    \item[a)] the human capability to differentiate between correct and incorrect AI advice and 
    \item[b)] to act upon that discrimination.
\end{itemize}
\end{theorem}
    
Now that we defined AR for our study, we derive a corresponding measurement. Current metrics in literature do not capture the ability to discriminate AI advice, including the final decision by the human. For example, the weight on advice (WOA) metric measures advice utilization \cite{sniezek2001trust}. This means the metric does not differentiate between correct or incorrect advice but instead measures the share of taken advice. Furthermore, performance metrics blur the effect of AR. For example, if AI has higher performance on a task than a human, blindly relying on AI without differentiating between correct and incorrect advice might increase the team performance. However, it will not result in the desired outcome that team performance exceeds the one of humans or AI conducting the task alone. For this reason, one cannot assume AR after a performance increase. This results in the need for a measurement concept that reflects how well humans are able to discriminate between correct and incorrect advice.

Following judge-advisor literature \cite{sniezek2001trust}, we propose to study AR in a sequential human-AI decision-making setup with two steps of human decision-making. 
\Cref{tab:metric} gives an overview of the different combinations based on a classification task. 
\begin{table}[b]
\centering
\small
\resizebox{\linewidth}{!}{%
\begin{tabular}{ccccc}
\hline
Initial human  & AI  & Relationship between initial human  & Human decision after  & Reliance \\
 decision & advice & decision and AI advice & receiving AI advice &
\\ \hline
Correct                & Correct   & Confirmation                                              & Correct                                  & n/a                    \\
Correct                & Correct   & Confirmation                                              & Incorrect                                & n/a                    \\
Incorrect              & Incorrect & Confirmation                                              & Correct                                  & n/a                    \\
Incorrect              & Incorrect & Confirmation                                              & Incorrect                                & n/a                    \\
Incorrect              & Correct   & Positive advice (PA)                     & Correct                                  & Positive AI reliance   \\
Incorrect              & Correct   & Positive advice (PA)                                                          & Incorrect                                & Negative self-reliance \\
Correct                & Incorrect & Negative advice (NA)                     & Correct                                  & Positive self-reliance \\
Correct                & Incorrect & Negative advice (NA)                                                           & Incorrect                                & Negative AI reliance   \\ \hline
\end{tabular}
}
\caption{Effect of AI advice on reliance.}
\label{tab:metric}
\end{table}
Note that for simplicity, we refer to classification problems. However, the measurement concept can be extended to regression problems as well. We consider a sequential decision process which can be described as follows: First, the human makes a decision, then receives AI advice. Second, the human is asked to update the initial decision, i.e., either adopt or overwrite the AI advice. 
This allows measuring AR in a fine-granular way. For example, the initial human decision can either be correct or incorrect in the classification setting. 
The follow-up AI advice can then either confirm the human's initial decision---or contradict it. We call the two contradictory cases positive and negative AI advice. For any further analysis, we propose to focus on the two contradictory cases as the confirmation cases do not allow to measure the human's discrimination ability and blur the actual discrimination capability. 
In general, if we do not consider the initial human decision, information about the human discrimination ability, including the consequent action, gets lost---it is not traceable how the human would have decided without the AI advice. Nevertheless, especially this interaction needs to be documented to research AR holistically. To give an example, imagine a case where the AI gives 10 times correct advice and 10 times incorrect advice. The human is initially 10 times correct. Now the question is, how are these 10 correct initial decisions distributed over the AI advice. If, for example, 8 correct initial decisions are followed by correct AI advice, this leads to a high accuracy after receiving correct AI advice. However, it is not distinguishable whether the resulting accuracy results from a good discrimination ability or other factors, e.g., uncertainty in one's own decision followed by confirmation by the AI prediction. 
By considering the initial human decision and focusing on the contradiction cases, the influence of confounding factors that can lead to misinterpretations can be minimized.

After receiving AI advice, humans need to decide whether to rely on AI or self-rely. We can now classify four types of reliance. First, \textit{positive AI-reliance}, which describes the case when the human is initially incorrect, receives correct advice and relies on that advice. Second, the case in which the human relies on the initial incorrect decision and neglects positive AI advice. This is denoted as \textit{negative self-reliance}. Third, if the human is initially correct and receives incorrect advice, this can either result in \textit{positive self-reliance}, i.e., neglecting the incorrect AI advice, or relying on it, which is denoted as \textit{negative AI-reliance}. We can now observe that the positive effect of AI-reliance and self-reliance depends on the quality of the AI advice. Therefore, we propose to measure AR on two dimensions.

On the first dimension, we calculate the ratio of cases where the human relies on correct AI advice under the condition that the decision was initially not correct, i.e., in which the human rightfully changes his mind to follow the AI advice.
\begin{equation}
    Relative\;positive\;AI\;reliance\;(RAIR) = \frac{Positive \;AI\; reliance}{Positive\;AI\;reliance + Negative\;self{\text-}reliance}
\end{equation}

On the second dimension, we propose to measure the relative amount of positive self-reliance in the presence of negative advice.
\begin{equation}
    Relative\;positive\;self{\text-}reliance\;(RSR) = \frac{Positive \;self{\text-}reliance}{Negative\;AI\;reliance + Positive\;self{\text-}reliance}
\end{equation}
\begin{figure}[h]
  \centering
  \includegraphics[width=0.45\linewidth]{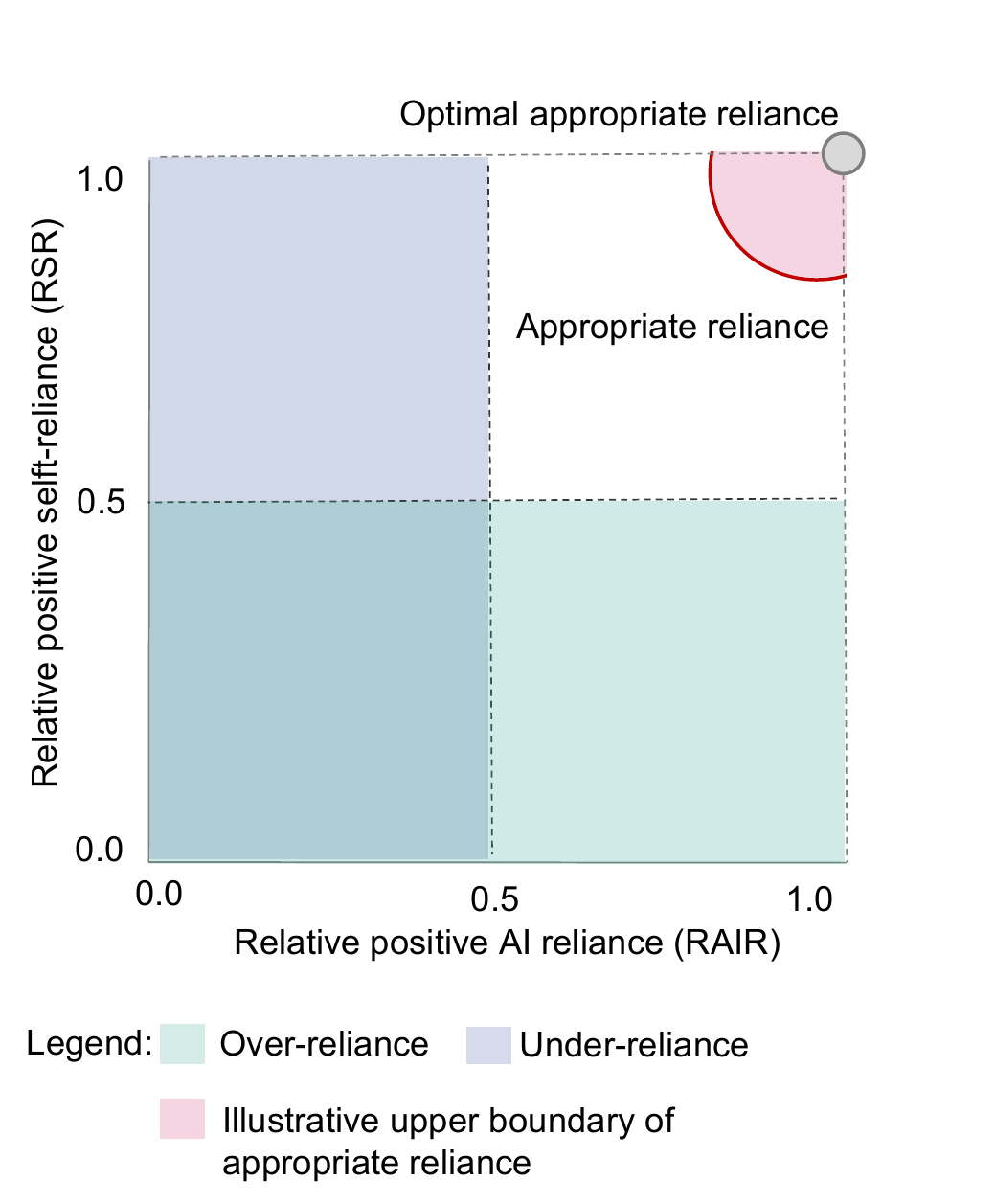}
  \caption{Two-dimensional depiction of appropriate reliance.}
  \label{fig:dimension}
\end{figure}

\Cref{fig:dimension} highlights both dimensions. On the x-axis, we depict the relative positive AI-reliance (\(RAIR\)), and on the y-axis, the relative positive self-reliance (\(RSR\)). 
The figure highlights the properties of the measurement concept. It ranges on both dimensions between 0 and 1. As a baseline, we can consider a random decision. 
The random baseline allows us to detect under- and over-reliance in a static case. If \(RSR\) and \(RAIR\) are below this threshold, we observe over- and under-reliance. More specifically, a \(RSR\) below the threshold means that a human performs worse than by chance in detecting incorrect AI advice, essentially representing an over-reliance on AI. Similarly, a \(RAIR\) below the random threshold means that a human differentiates correct AI advice worse than a random guess, essentially representing an under-reliance on AI. We depict the decision-making threshold in \Cref{fig:dimension} to illustrate this reasoning. Furthermore, we can use the space to analyze the effect of experimental treatments. For example, a treatment that increases \(RAIR\) and decreases \(RSR\) actually just increases over-reliance on AI. Similarly, a condition that increases \(RSR\) while decreasing \(RAIR\) points towards under-reliance.

We refer to the theoretical goal of having a \(RSR\) and a \(RAIR\) metric of ``1'' as optimal AR. Most likely, this theoretical goal will not be reached in any practical context as humans will not always be able to perfectly discriminate on a case-by-case basis whether they should rely on AI advice. Furthermore, random errors will reduce AR as they cannot be discriminated against. Therefore, optimal AR will most likely be a theoretical goal. Lastly, the area with \(RAIR\) and \(RSR\) larger than the random threshold encompasses the proportion of final decisions that did not occur by chance. Therefore, AR is defined as every combination larger than the random threshold. In \Cref{fig:dimension} it refers to the right top quadrant. This means AR is not binary but a tuple of \(RAIR\) and \(RSR\) above the random threshold with the theoretical optimum of ``1''.

To illustrate our measurement concept, in the following \Cref{sec:illustration}, we describe the results of an experimental study.






\section{Illustration of Appropriate Reliance on AI Advice}\label{sec:illustration}

To illustrate the proposed measurement concept, we conducted a user study. The study's goal is to highlight how our measurement concept can be used to evaluate human-AI decision-making experiments with regard to AR. For illustration, we focus on the explainability of AI advice as a design decision. XAI is intensively discussed in research with regards to its impact on human-AI decision-making in general and AR in specific \cite{bansal2021does,buccinca2021trust,lai2021towards,smith2020no}.

To discriminate advice, humans need information that can approximate the quality of advice, e.g., the uncertainty of the advisor or explanations \cite{sniezek2001trust}. The emerging research stream of XAI \cite{hoffman2018metrics} aims to equip human users with such insight into AI advice. Explanations might help the human decision-maker better judge the quality of an AI-based decision. An analogy would be the interaction between a consultant, who provides advice, and his client. To assess the quality of the advice, the client will ask the consultant to describe the reasoning. Based on the explanations, the decision-maker should be able to determine whether the advice can be relied upon or not. The same logic should hold for an AI advisor.

On the other hand, experimental studies indicate that explanations in human-AI decision-making can lead to over-reliance \cite{bansal2021does,zhang2020effect}. Research shows that explanations are sometimes interpreted more as a general sign of competence \cite{buccinca2021trust} and have a persuasive character \cite{bussone2015role}. This is supported by psychology literature which has shown that human explanations cause humans to agree even when the explanation is wrong \cite{koehler1991explanation}. The same effect could occur when the explanations are generated by AI. Therefore, there exists an ambiguous trade-off between enabling the human to discriminate the AI’s advice and, on the other hand, the tendency that the sole existence of an explanation could increase AI reliance \cite{bansal2021does}. 
In this illustrative study, we use this often discussed ambiguity to highlight the advantages of our measurement concept.

As an experimental task, we have chosen a deceptive hotel review classification. Humans have to differentiate whether a given hotel review is deceptive or genuine. \citet{ott2011finding,ott2013negative} provide the research community with a data set of 400 deceptive and 400 genuine hotel reviews. The deceptive ones were created by crowd-workers, resulting in corresponding ground truth labels.

\begin{figure}[h]
  \centering
  \includegraphics[width=0.9\linewidth]{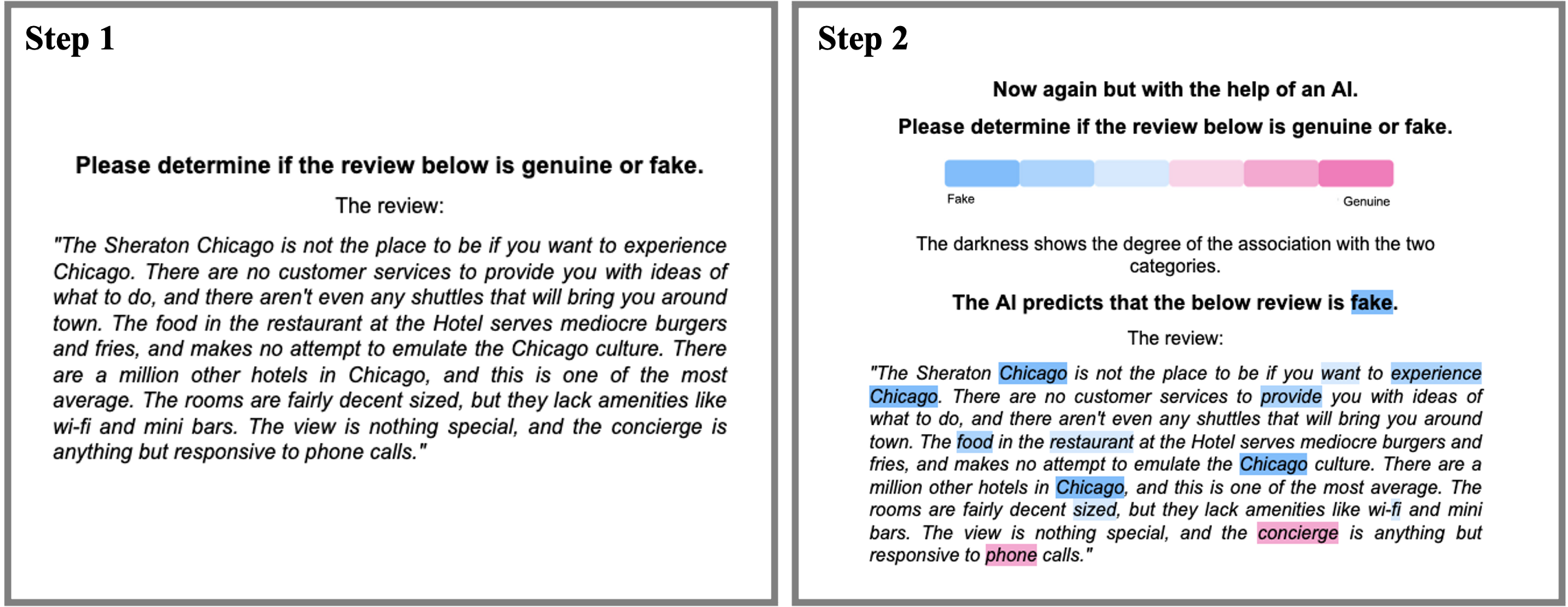}
  \caption{Online experiment graphical user interface for the XAI treatment. The ground truth of the exemplarily shown hotel review is ``fake''. The design of the interface is adapted from \citet{lai2020chicago}.}
  \label{fig:experiment}
\end{figure}

The implemented AI is based on a Support Vector Machine with an accuracy of 86\%, which is a performance that is similar to the performance in related literature \cite{lai2020chicago}. For the XAI condition, we use a state-of-the-art explanation technique, namely LIME feature importance explanations \cite{ribeiro2016model}. Feature importance aims to explain the influence of an independent variable on the AI's decision in the form of a numerical value. Since we deal with text data, a common technique to display the values is to highlight the respective words according to their computed influence on the AI's decision \cite{lai2020chicago}. We additionally provide information on the direction of the effect and differentiate the values into three effect sizes following the implementation of \citet{lai2020chicago} (see step 2 in \Cref{fig:experiment}).

For the AR measurement concept, a sequential task processing is essential. In our study, this means the human first receives a review without any AI advice, i.e., just the plain text, and classifies whether the review is deceptive or genuine (see step 1 in \Cref{fig:experiment}). Following that, the human either receives a simple AI advice statement, e.g. ``the AI predicts that the review is fake'' or the AI advice and additional explanations (see step 2 in \Cref{fig:experiment}). This sequential two-step decision-making allows us to measure AR.

The participants were recruited using the platform Prolific.co. In total, we conducted the experiment with 200 participants. In each treatment, participants were provided with 16 reviews---8 correct and 8 incorrect ones.

\begin{figure}[h]
  \centering
  \includegraphics[width=0.45\linewidth]{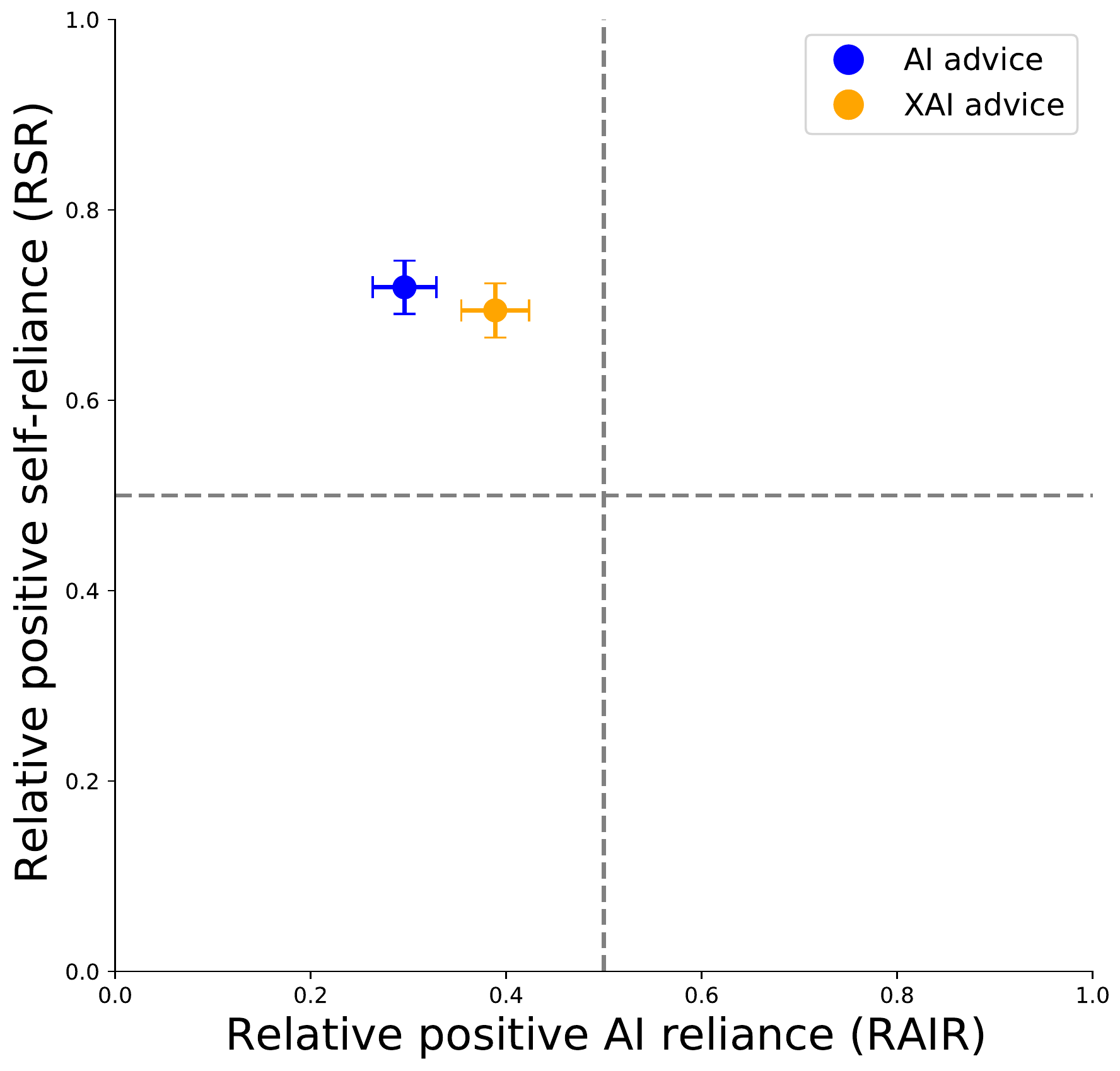}
  \caption{Illustration of appropriate reliance on AI advice including standard errors.}
  \label{fig:illustration}
\end{figure}

We depict the results of the experiment in \Cref{fig:illustration}. They highlight in the AI condition a high \(RSR\) of $0.72$ ($\pm0.03$) and a relatively low \(RAIR\) of $0.3$ ($\pm0.03$). 
This indicates that humans in the setting were able to differentiate wrong AI advice and self-rely to a high degree. The \(RAIR\) of $0.3$ shows that we can observe under-reliance on AI as the \(RAIR\) is below the random guess of our binary classification task. Our experiment thereby highlights the general tendency of humans to ignore AI advice that is in literature usually discussed as algorithm aversion \cite{dietvorst2015algorithm}.

In the XAI condition, we can observe a significant increase ($t=-1.95$, $p =0.05$) in \(RAIR\) from $0.3$ ($\pm0.03$) to $0.39$ ($\pm0.03$) while the \(RSR\) does not change significantly ($0.7\pm0.03$ for AI and $0.72\pm0.03$ for XAI, $t = 0.61$, $p = 0.54$). This means explanations of AI decisions can reduce under-reliance. It is important to highlight that under-reliance is not reduced simply by relying more often on AI advice, as this would have also reduced the \(RSR\) significantly. Thus, our experiment indicates that explanations may have a positive effect on human-AI decision-making and do not necessarily result in over-reliance.
Our user study further illustrates the usage of our measurement concept and the potential to apply it in experimental studies.

\section{Discussion}\label{sec:discussion}
In this paper, we conducted a review on AR and proposed a measurement concept for its measurement in human-AI decision-making. Subsequently, we illustrated this concept in the scope of a user study to highlight its capability. With our approach, AR can be measured on a more fine-granular level. Thus, it can be leveraged in future experimental studies to address questions on how to design for AR. 

In this context, research needs to investigate the capability of AR to discriminate between incorrect and correct AI advice and evaluate possible impact factors. For example, the potential to discriminate might be different between \(RSR\) and \(RAIR\). One would assume that it might be easier to discriminate negative advice than positive advice, as in the negative advice condition, the human is initially, per definition, able to solve the task. In contrast, it might be challenging to discriminate positive AI advice after failing to solve a task alone correctly. Our initial illustrative experiment also showed these differences. The experiment further highlighted that while XAI might address \(RAIR\), it does not seem to influence \(RSR\). Therefore, research needs to investigate the differences in detail and find proper ways to address them. 

Furthermore, researchers need to investigate factors beyond the discrimination capability that influence the behavioral part of AR. For instance, research models could incorporate attitudes as well as human biases that might influence AR. Among others, important constructs that have been evaluated in the judge-advisor literature concerning advice utilization are human confidence \cite{long2020ai}, perception \cite{schoeffer2021perceptions} and trust \cite{sniezek2001trust}. 
Other research has shown the influence of human bias on AR. For example, so-called egocentric discounting---humans systematically overweighting their own decisions---increases under-reliance on AI \cite{yaniv2000advice}. Some of these attitudes might enable enhanced discrimination, such as engagement in AI advice \cite{hoffman2018metrics}. Others could potentially harm AR. For example, maximizing trust in AI could lead to ``blind trust'' and consequently lead to a situation where humans accept all advice \cite{bansal2021does}. Similar phenomena could happen in terms of cognitive constraints. Research in automation has shown that humans tend always to follow the path of least cognitive effort \cite{skitka1999does}. Simply accepting could therefore be a preferred human decision. Research needs to investigate these factors in future work.

Lastly, we want to emphasize several limitations of the proposed measurement concept. First, the concept is limited to classification tasks but will be extended in future work. First approaches can be found in the work of \citet{petropoulos2016big}.
Furthermore, the sequential task setup necessary for our measurement concept has some disadvantages as it changes the task itself. Since conducting the same task initially alone before receiving AI advice, the human is already mentally prepared and might react differently than after directly receiving AI advice. Moreover, sequentially conducted tasks with AI advice might not always be possible or desired in real-world cases. Therefore, the measurement should be seen as an approximation of real human behavior. Instead of having a sequential task setup, one alternative option could be to simulate a human model based on a data set of task instances solved by humans without AI advice. This simulation model could approximate the initial human decision within a non-sequential task setting. However, also this approach is an approximation of real human behavior. Future work should compare both approaches.

\section{Conclusion}\label{sec:conclusion}
Many researchers highlighted the need for humans to rely on AI advice appropriately, i.e., being able to discriminate AI advice quality and acting upon it for the best possible human-AI decision-making \cite{buccinca2021trust,bansal2021does,zhang2020effect,lu2021human}. However, current research is missing a measurement concept for AR that allows the evaluation of human-AI decision-making experiments.

Therefore, in this article, we develop a new measurement concept for quantifying AR in the human-AI decision-making context. Specifically, we propose to view AR as a two-dimensional construct that measures the capability to discriminate the quality of advice and behave accordingly. The first dimension considers the relative positive effect of relying on AI advice, whereas the second dimension assesses the relative positive self-reliance in the presence of incorrect AI advice. Subsequently, we illustrate our measurement concept and provide an outlook on future research. Our research provides a basis for future studies to evaluate the impact factors of AR and develop designs possibilities to improve human-AI decision-making.

\bibliographystyle{ACM-Reference-Format}
\bibliography{sample-base}


\end{document}